\documentclass[aps,pra, twocolumn]{revtex4}
\usepackage{graphicx}
\usepackage{amssymb}
\usepackage{amsfonts}
\usepackage{amsmath}
\usepackage{lmodern}
\usepackage{mathrsfs}
\usepackage{mathdots}
\usepackage{collref}
\usepackage[colorlinks=true, citecolor=blue, urlcolor=blue]{hyperref}
\usepackage{tikz}
\usepackage{bm}
\begin{document}
\title{Effects of local decoherence on quantum critical metrology}
\author{Chong Chen}\email{chongchenn@gmail.com}
 \affiliation{Department of Physics, Centre for Quantum Coherence, and The Hong Kong Institute of Quantum Information Science and Technology, The Chinese University of Hong Kong, Shatin, New Territories, Hong Kong, China}
 \author{Ping Wang}
 \affiliation{Department of Physics, Centre for Quantum Coherence, and The Hong Kong Institute of Quantum Information Science and Technology, The Chinese University of Hong Kong, Shatin, New Territories, Hong Kong, China}
\author{Ren-Bao Liu}
 \affiliation{Department of Physics, Centre for Quantum Coherence, and The Hong Kong Institute of Quantum Information Science and Technology, The Chinese University of Hong Kong, Shatin, New Territories, Hong Kong, China}
\date{\today}

\begin{abstract}
  The diverging responses to parameter variations of systems at quantum critical points motivate schemes of quantum critical metrology (QCM), which feature sub-Heisenberg scaling of the sensitivity with the system size (e.g., the number of particles). This sensitivity enhancement by quantum criticality is rooted in the formation of Schr\"{o}dinger cat states, or macroscopic superposition states at the quantum critical points. The cat states, however, are fragile to decoherence caused by coupling to local environments, since the local decoherence of any particle would cause the collapse of the whole cat state. Therefore, it is unclear whether the sub-Heisenberg scaling of QCM is robust against the local decoherence. Here we study the effects of local decoherence on QCM, using a one-dimensional transverse-field Ising model as a representative example. We find that the standard quantum limit is recovered by single-particle decoherence. Using renormalization group analysis, we demonstrate that the noise effects on QCM is general and applicable to  many universality classes of quantum phase transitions whose low-energy excitations are described by a $\phi^{4}$ effective field theory. Since in general the many-body entanglement of the ground states at critical points, which is the basis of QCM, is fragile to quantum measurement by local environments, we conjecture that the recovery of the standard quantum limit by local decoherence is universal for QCM using phase transitions induced by the formation of long-range order. This work demonstrates the importance of protecting macroscopic quantum coherence for quantum sensing based on critical behaviors.
\end{abstract}
\maketitle
\section{Introduction}
Quantum metrology distinguishes parameters using the distinguishability of quantum states \cite{Giovannetti2004,Giovannetti2011}. Quantum states that have suppressed fluctuations of certain observables (such as squeezed states and globally entangled states) and quantum evolutions that are sensitive to parameter changes can enhance the sensitivity of parameter estimation \cite{Wineland1992,Strobel2014,Pezze2018}. An $N$-body entanglement state, e.g., a GHZ state, can realize a sensitivity with the Heisenberg limit scaling $1/N$, which is far beyond the standard quantum limit scaling $1/\sqrt{N}$ for a product state \cite{Giovannetti2006}. 
Systems around quantum critical points have many-body entanglement \cite{Sachdev2007,Vidal2004} and their evolutions have high susceptibility to external fields \cite{Quan2006,Chen2013}. These features motivate schemes of using quantum critical systems for parameter estimation \cite{Zanardi2008,Invernizzi2008,Skotiniotis2015,Frerot2018}, which are known as quantum critical metrology (QCM). Approaching to the critical point, the ground-state fidelity susceptibility \cite{Invernizzi2008} presents a super-Heisenberg scaling with the number of particles ($N^\alpha$, $\alpha>1$), which alludes to a sensitivity beyond the Heisenberg limit $N^{-1}$ \cite{Gong2008,Gu2008,Greschner2013}. Further analysis considering the time consumption of the evolution shows that the parameter sensitivity has actually a sub-Heisenberg scaling ($N^{-\alpha}$, $1/2<\alpha<1$) , lying between the standard quantum limit and the Heisenberg limit \cite{Rams2018,Pezze2019}, which nonetheless still represents a significant enhancement of sensitivity.

Decoherence due to coupling to environments, however, may reduce or even eliminate the sensitivity enhancement by many-body entanglement. For a finite-$N$ non-interacting system prepared in a macroscopic superposition state (a Schr\"{o}dinger cat state), the environmental noise sets the best possible scaling between the standard quantum limit and the Heisenberg limit \cite{Dorner2009,DAmbrosio2013}; even worse in the thermodynamic limit $N\rightarrow \infty$, an infinitesimal noise can reduce the scaling from the Heisenberg limit to the standard quantum limit \cite{Fujiwara2008,Escher2011,Demkowicz-Dobrzanski2012,Yuan2017}. The QCM, different from the entanglement-based metrology using interaction-free systems \cite{Giovannetti2004,Giovannetti2006,Escher2011}, takes advantage of the fact that both quantum state and evolution in sensing process are based on the same many-body system \cite{Frerot2018,Rams2018, Garbe2020}.  A natural question for QCM is: How would the noise affect the sensitivity scaling?  One clue is that the quantum entanglement of the ground state at the critical point due to the formation of a long-range order is fragile to local measurements by environments \cite{Huelga1997} (which is the underlying mechanism of spontaneous symmetry breaking \cite{Beekman2019}). Recently an intriguing study on the  $p$-body Markovian dephasing dynamics of $N$-spin GHZ states evolving under a $k$-body Hamiltonian shows an $N^{-(k-p/2)}$ scaling of the estimation error \cite{Beau2017}.

Here we study the effects of local decoherence on QCM, focusing on the scaling of sensitivity with the number of particles.  We first consider the one-dimensional transverse-field Ising model as a representative example. This model has exact solutions and has been used to show that the sensitivity of parameter estimation is enhanced by the many-body entanglement at the critical point \cite{Zanardi2008,Invernizzi2008,Skotiniotis2015,Frerot2018}. To investigate the effects of local decoherence, we couple the spins to local bosonic environments. The coupling to local boson baths dramatically modifies the phase diagram of the model \cite{Werner2005}. Consistent with the modified phase diagram, we find that the sub-Heisenberg scaling of the sensitivity is reduced to the standard quantum limit. Then, using the universal scaling law established in Ref. \cite{Pankov2004}  by renormalization group analysis, we demonstrate that the criticality-enhancement of sensitivity being suppressed by local decoherence is applicable to many universality classes of quantum phase transitions whose low-energy excitations are described by a $\phi^4$ effective field theory.

\section{QCM and  application to Ising chains}
Consider a parameter $J$ of a system with a quantum critical point $J_{C}$. We assume the quantum phase transition is caused by the formation of a long-range order characterized by a local order parameter $M$.  Near the critical point, the correlation length diverges as $\xi\sim|J-J_C|^{-\nu}$ and therefore becomes the only relevant scale. According to the scaling hypothesis, thermodynamic quantities scale by power laws with the correlation length. Examples are the order parameter $M \sim\xi^{-\beta/\nu}$ and the susceptibility $\chi=\frac{\partial M}{\partial h}\sim\xi^{\gamma/\nu}$, where $h$ is an external field coupled to the order parameter. The diverging susceptibility at the critical point is the basis of QCM for estimating the parameter $h$ \cite{Rams2018}. First, the system  is prepared at the ground state at quantum critical point; then, a small field $h$ is applied; finally, after the free evolution of time $t$, a quantum measurement is performed and the result is compared with that obtained without applying the field $h$. The sensitivity is defined as the smallest  $h$ that yields measurement difference greater than the quantum fluctuation for an evolution time $t$, i.e., $\eta_{h} =h_{\rm{min}} \sqrt{t}$. The sensitivity has a theoretical lower bound, known as the Cram\'{e}r-Rao bound \cite{Braunstein1994} $\eta_{h} \ge {1\over\sqrt{F(h)/t}}$, where the quantum Fisher information $F\left(h\right)$ is related to the spectral function $\chi^{\prime\prime}\left(q=0,\omega\right)=\pi N\sum_{n\neq0}{|\langle0|{{\hat{M}}}| n\rangle|^2\delta(\omega-E_n)}$ by \cite{Hauke2016}
\begin{equation}
	F(h)=\frac{8N}{\pi} \int d\omega \chi^{\prime\prime}(q=0,\omega) \frac{1-\cos(\omega t)}{\omega^{2}}.
	\label{eq:QFI}
\end{equation}
Here $|0\rangle$ and $|n\rangle$ are the ground state and the $n$-th excited state, respectively, and $E_{n}$ is the excitation energy.

The critical behaviors are determined by the low-energy excitations. Around the critical point, the gap of low-energy excitations scales with the correlation length by $\Delta \sim\xi^{-z}$,where $z$ is the dynamic critical exponent. Applying this scaling equation to Eq. \eqref{eq:QFI} and using the fluctuation-dissipation theorem $\chi=\int{\frac{d\omega}{\pi\omega}\chi^{\prime\prime}\left(q=0,\omega\right)}\sim\xi^{\gamma/\nu}$, we get \cite{Rams2018}
\begin{subequations}
\begin{eqnarray}
    \label{eq:QFI-Scaling-a}
	F(h)&\sim& N t^{2} \xi^{\gamma/\nu-z}, ~ ~\mathrm{for}~ ~ t< \xi^{z}, \\
    \label{eq:QFI-Scaling-b}
	F(h)&\sim& N \xi^{\gamma/\nu+z}, ~ ~ ~ ~\mathrm{for}~~ t \ge \xi^{z}.
\end{eqnarray}
	\label{eq:QFI-Scaling}
\end{subequations}

Specifically, we consider a one-dimensional transverse-field Ising model with the Hamiltonian
\begin{equation}
	\hat{H}_{0}=-J \sum^{N}_{i=1} \hat{\sigma}^{z}_{i} \hat{\sigma}^{z}_{i+1}-B \sum^{N}_{i=1} \hat{\sigma}^{x}_{i},
	\label{eq:IsingModel}
\end{equation}
where $\hat{\sigma}_{i}^{x/y/z}$ are the Pauli matrices of the $i$-th spin along the $x/y/z$ directions. The order parameter operator is $\hat{M}\equiv \frac{1}{N} \sum^{N}_{i=1} \hat{\sigma}_{z}$. The parameter to be estimated couples to the spins by $h \sum^{N}_{i}\hat{\sigma}_{z}$. This model has exact solution and the critical point and the critical exponents are derived as $J_{C}=|B|$, $\gamma=7/\ 4$, $\nu=1$, and $z=1$ \cite{Pfeuty1970}. Applying these exponents to Eq. \eqref{eq:QFI-Scaling-b}, we obtain a super-Heisenberg quantum Fisher information scaling \cite{CamposVenuti2007} $F(h)\sim N^{15/4}$
at the long-time limit ($t>N$), where the condition $\xi \sim N$ for the one-dimensional system has been used. The super-Heisenberg scaling comes from the fact that the evolution time is absorbed into the scaling of $N$ \cite{Rams2018}.  If the evolution time $t<\xi^z$, the scaling of the quantum Fisher information becomes $F(h) \sim t^{2} N^{7\over 4}$, which yields a sub-Heisenberg limit $\eta_{h} \sim t^{-\frac{1}{2}} N^{-\frac{7}{8}}$ \cite{Rams2018}.

To relate the sensitivity enhancement to many-body entanglement, we use the average variance $N_{\mathrm{e}}\equiv N  \mathrm{Var}(M)$ to characterize the multipartite entanglement \cite{Hauke2016}, which means the number of neighboring spins that are entangled. The average variance is related to the two-site entanglement, which can be used to characterize quantum phase transitions \cite{Amico2008}. The Fisher information can be written as $F\left(h\right) \approx4Nt^2N_\mathrm{e}$.  As proved in Refs. \cite{Hyllus2012,Toth2012}, $F\left(h\right)/\ (4t^2)\le Nk$ if the state is $k$-producible (i.e., there are at most $k$ neighboring  particles entangled), so the critical ground state has at least $N_\mathrm{e}\sim N^{3/4}$ neighboring particles entangled. The fact that $N_\mathrm{e} \rightarrow O\left(N^\alpha\right)$ with $\alpha>0$ indicates long-range correlation and hence long-range order.

 \section{Effects of local decoherence}
 We couple each spin of the Ising chain in Eq. \eqref{eq:IsingModel} to an independent bosonic environment \cite{Werner2005}. The total Hamiltonian reads
 \begin{equation}
	\hat{H}=\hat{H}_{0}+\sum_{i,k} [\hat{\sigma}^{z}_{i} g_{k}(\hat{b}_{i,k}+\hat{b}^{\dagger}_{i,k})+\omega_{k}\hat{b}^{\dagger}_{i,k} \hat{b}_{i,k}],
	\label{eq:noisyH}
\end{equation}
where $\hat{b}_{i,k}$ ($\hat{b}^\dagger_{i,k}$) is the annihilation (creation) operator of the $k$-th mode  of the bosonic bath coupled to the $i$-th spin, with frequency $\omega_k$ and coupling strength $g_k$ (both assumed site independent). The noise spectrum, $J(\omega)=\sum_{k}{g_k^2\delta(\omega-\omega_k)}$, is the same in all sites by assumption. We set the environment temperature to be zero and take the noise spectrum as Ohmic, i.e., $J(\omega)=\alpha\omega e^{-\omega/\omega_\mathrm{c}}$, with a cutoff $\omega_\mathrm{c}$ and a dimensionless coupling constant $\alpha$. The effects of other types of noise spectra are discussed later. Unlike the studies based on non-equilibrium steady state transitions \cite{Wald2020,Garbe2020}, here the formulation of critical quantum metrology in equilibrium quantum transitions (via adiabatic switching-on of couplings) allows a general critical scaling analysis.

The local decoherence can be understood as quantum measurement ``performed" by the local bosonic environments. When the system state is such that a spin has expectation value $\left\langle{\hat{\sigma}}_i^z\right\rangle=\pm1$, the boson mode $k$ gets a $\pm g_k$ displacement. Thus, the boson modes measure the spin ${\hat{\sigma}}_i^z$ and the spin collapses to one of the basis states. Due to the long-range entanglement, the whole spin chain collapses into a state robust against local decoherence (essentially a product state). Such a state collapse process is the mechanism of spontaneous symmetry breaking. The parameter estimation via a product state has a sensitivity scaling in the standard quantum limit $1/\sqrt{N}$.

To quantitatively study the decoherence effects, we map the one-dimensional quantum transverse-field Ising model to a two-dimensional classical Ising model using Suzuki-Trotter decomposition with discrete imaginary time $\tau=1,\cdots,N_\tau$ \cite{Suzuki1993}, which is exact for $N_{\tau}\rightarrow \infty$. After integration of the bosonic modes, an effective Ising action is obtained as
\begin{align}
  S_{\mathrm{eff}}= &-\sum^{N_{x}}_{i=1} \left[\sum^{N_{\tau}}_{\tau=1} (J s_{i,\tau} s_{i+1,\tau}+\Gamma s_{i,\tau} s_{i,\tau+1}) \right. \nonumber \\
  & +\left. \frac{\alpha}{2}\sum^{N_{\tau}}_{\tau<\tau^{\prime}} \frac{(\pi/N_{\tau})^2s_{i,\tau} s_{i,\tau^{\prime}}}{\sin^{2}\frac{\pi(\tau^{\prime}-\tau)}{N_{\tau}}} \right],
  \label{eq:IsingAction}
\end{align}
where $s_{i,\tau}=\pm1$ are classical spins and $\Gamma=-\frac{1}{2}\ln{(\tanh{B})}$ with the lattice constant along the $\tau$ direction taken as unity. There emerges a long-range effective interaction along the imaginary time direction.

We reproduce the phase diagram and the critical exponents presented in Ref. \cite{Werner2005}, as shown in \cite{SM}. With the transverse field $B$ fixed, the coupling to the local environments extends the phase boundary from a critical point to a critical line in the $\alpha-J$ plane. When $J>J_C$, the system is always in the ferromagnetic phase as expected. When $J\le J_C$, a transition between the paramagnetic and the ferromagnetic phases occurs at a critical noise strength $\alpha_C$, which increases with decreasing $J$.

The destruction of the long-range entanglement by the local decoherence is evidenced by the decay of the spin correlation $C(r,0)=\langle \hat{\sigma}^{z}_{i+r} \hat{\sigma}^{z}_{i}\rangle-\langle \hat{\sigma}^{z}_{i} \rangle^2$ (with $\hat{\sigma}^{z}_{i} \mapsto s_{i,0}$ in action \eqref{eq:IsingAction}). Around the critical point, the scaling theory gives $C(r\gg 1, 0)\sim r^{z+\eta-1}$, since $C(r,0)=\int \frac{d\omega}{2\pi} \int \frac{dk}{2\pi} \tilde{C}(k,\omega) e^{i kr}$ and $\tilde{C}(k,\omega)\sim (\omega^{2/z}+k^{2})^{1-\eta/2}$ at the critical point \cite{Werner2005}. The numerical fitting $C(r\gg 1,0)=a r^{-b}+c$ yields the critical exponent $z+\eta-1=0.25(2)$ for the quantum critical point without decoherence and $z+\eta-1=1.0(2)$ for the critical point with decoherence. The faster decay of the correlation $C(r,0)$ in the decoherence case indicates that the coupling to the local environments destroys the multipartite entanglement. Indeed, the average variance $N_\mathrm{e}\approx\int_{0}^{N}{dr\ C(r,0)}$  changes from the power law scaling $N_\mathrm{e}\sim N^{3/4}$ to the logarithm scaling $N_\mathrm{e}\sim\log{N}$.

The quantum Fisher information, in term of the spin correlation,
\begin{equation}\label{eq:QFI-Corr}
  F\left(h\right)\approx Nt^2\int dr C(r,0),
\end{equation}
is reduced by decoherence to
\begin{equation}
F_{\rm{noisy}}\left(h\right)\sim Nt^2,
\label{eq:QFI-noisy}	
\end{equation}
up to a logarithm modification $\log{N}$, at the critical point $J<J_{C}$ and $\alpha=\alpha_{C}$. Equation \eqref{eq:QFI-noisy} is consistent with the scaling analysis in Eq. \eqref{eq:QFI-Scaling-a} where $F\left(h\right)\sim Nt^2\xi^{\gamma/\nu-z}\sim Nt^2\xi^{2-\eta-z}\sim Nt^2$. Here we have used the fact that $z+\eta\approx2$ and the Fisher scaling law $\gamma/\nu=2-\eta$.  In conclusion, the local decoherence recovers the standard quantum limit.

\section{Effects of non-Ohmic noises}
Above we have assumed that the noise spectrum has the Ohmic form in Eq. \eqref{eq:noisyH}, where the scaling law $z+\eta=2$ holds. Here we consider a more general power-law noise spectrum $J\left(\omega\right)=\alpha\omega^s\omega_\mathrm{c}^{1-s}e^{-\omega/\omega_\mathrm{c}}$ with a cutoff $\omega_\mathrm{c}$. This spectrum has density $\sim\omega^s$ at the low-energy limit. The effective Ising model, after integration of the bosonic modes as in Eq. \eqref{eq:IsingAction}, has a long-range interaction $\sim\left(\tau-\tau^\prime\right)^{-(1+s)}$ in the imaginary time dimension. The previous studies of the long-range Ising model show that the critical exponents have three regimes depending on the value of $s$: the mean-field regime $0<s<2/3$ where $z=2/s$ and $\eta=0$, the continuous regime $2/3 \le s<2$ where $z+\eta$ varies continuously from $3$ to $5/4$, and the Ising universality regime $s\geq 2$ where $z=1$ and $\eta=1/4$ \cite{Defenu2017,Fey2016,Zhu2018,Sperstad2012}. The Fisher information $F\left(h\right)\sim Nt^2\xi^{2-\eta(s)-z(s)}$ is a monotonically decreasing function of $s$. The standard quantum limit $F\left(h\right)\sim Nt^2$ is reached at the threshold $s=1$ where $z+\eta=2$. For $s>1$, there is always enhancement by the quantum criticality.  In particular, the noises become irrelevant when $s\geq2$ where $z=1$ and  $\eta=1/4$ are constants. Physically, a bosonic bath with $s\geq2$ is essentially a gapped system and therefore has no effects on the spin decoherence at zero temperature.

On the other side of the threshold, $s<1$, the local decoherene can even reduce the Fisher information to a sub-standard quantum limit,
$F\left(h\right)\sim N^{1-x}t^2$ with $x>0$.  Physically, this is because the strong damping of the spins by the sub-Ohmic noise at low frequencies would make the spin dynamics insensitive to the field  $h$ (similar to the case of over-damped oscillators). Near the critical point, the correlation between the spins enhances the over-damping and therefore leads to the sub-standard quantum limit scaling of sensitivity.

\section{Universal decoherence effects on QCM}
Now we demonstrate that the effects of local decoherence on the QCM is universal, using the renormalization group analysis in Ref. \cite{Pankov2004}.

The low-energy excitations of a broad range of quantum critical systems can be captured by a $\phi^{4}$ effective theory described by the action
\begin{align}
  S=&\iint d x^{D}  d\tau \left[\frac{1}{2}(\bm{\nabla}_{x} \phi(\mathbf{x},\tau))^{2}+\frac{1}{2}(\partial_{\tau} \phi(\mathbf{x},\tau))^{2} +\frac{\tilde{\Delta}}{2} \phi^{2}(\mathbf{x},\tau) \right.\nonumber \\
  &\left. - A \int d\tau^{\prime} \frac{\phi(\mathbf{x},\tau)\phi(\mathbf{x},\tau^{\prime})}{2\pi(\tau-\tau^{\prime})^{2}} +\frac{\mu_{0}}{4!} \phi^{4}(\mathbf{x},\tau) \right],
  \label{eq:FieldTheory}
\end{align}
where $\phi(\mathbf{x},\tau)$ is the ordering field, $D$ is spatial dimension ($D=1$ for the Ising chain in Eq. \eqref{eq:IsingAction}), the coefficients before the $\bm{\nabla}_{x} \phi\left(\mathbf{x},\tau\right)$ and $\partial_\tau\phi\left(\mathbf{x},\tau\right)$ terms are absorbed into the definitions of $x$ and $\tau$, $A\propto\alpha$ is the noise strength,  the gap $\tilde{\Delta}$ and the interaction strength $\mu_0$ are phenomenological parameters that can be determined by fitting to experimental or numerical data. With the $\phi^4$ term neglected, the free propagator is $C_0^{-1}\left(\mathbf{q},\omega\right)=(\tilde{\Delta}+A\left|\omega\right|+\omega^2+q^2)$. From the free propagator one can see a crossover of the dynamical critical exponent between $z=1$ and $z=2$ with lowering the energy scale at the critical point where $\tilde{\Delta}=0$. When $\omega\gg A$, the resonance has a linear dispersion $\omega^2\sim q^2$ (i.e., $z=1$); when $\omega\ll A$, the dispersion becomes $\omega\sim q^2/A$ (i.e., $z=2$). At the critical point, the system behaviors are dominated by the low energy excitation with divergent wavelength ($q\rightarrow 0$), therefore $z=2$.

When the $\phi^4$ interaction is taken into consideration, the dimension analysis shows that $[\phi\left(x,t\right)]\sim \xi^\frac{2-D-z}{2}$ \cite{Pankov2004}. Such an analysis yields an upper critical dimension $D=2$ with the assumption that $z=2$. When $D \geq 2$, e.g., for a  2D transverse-field Ising model, $\left[\int d x^D d\tau\phi^4\right]=\xi^{4-D-z}\sim O\left(1\right)$  and the mean-field theory that neglects the $\phi^4$ fluctuation becomes exact. Consequently $z=2$ and $\eta=0$ hold for $D \geq 2$. Below the upper-critical dimension an $\epsilon=2-D$ expansion method can be used to analyze the effect of the $\phi^4$ term. The analysis \cite{SM} shows that the susceptibility has a universal expression $\tilde{C}\left(\mathbf{q},i\omega\right)=q^{-2+\eta}\phi(\frac{\omega}{cq^{2-\eta}})$, which leads to the universal scaling relation
$[\omega]\sim\xi^{2-\eta}$ and therefore the scaling law for the quantum phase transition with the decoherence effects \cite{Pankov2004}
\begin{equation}
  z+\eta=2.
  \label{eq:scalinglaw}
\end{equation}
This scaling law indicates that the equal imaginary time correlation $C\left(r,0\right)\sim r^{-d}$ and hence $F(h)\sim N t^{2}$. Therefore, the standard quantum limit is restored for noisy quantum critical systems that satisfy the scaling law \eqref{eq:scalinglaw}.

The quantum-to-classical mapping in Eq. \eqref{eq:IsingAction} and the $\phi^{4}$ theory \eqref{eq:FieldTheory}, according to Refs. \cite{Hertz1976,Millis1993}, are correct for generic quantum phase transitions as long as only the low-energy excitations (i.e., low-temperature physics) are considered. The scaling law in Eq. \eqref{eq:scalinglaw} and hence the conclusion that the local decoherence recovers the standard quantum limit of QCM hold if the effective $\phi^{4}$ action without the noise term (i) has a linear dispersion at the critical point and (ii) has a real value (without an imaginary term, e.g., a topological $\theta$-term). Note that the renormalization group analysis applies also to multi-component $\phi^{4}$ theory \cite{Pankov2004}. Therefore, the recovery of the standard quantum limit of QCM by local decoherence is the case for a broad range of universality classes of quantum phase transitions. Examples include the superfluid-insulator transitions of Bose-Hubburd models and the Neel transitions of antiferromagnetic Heisenberg models in different dimensions.

We conjecture that the conclusion may even hold more generally for all quantum phase transitions that involve the formation of long-range orders. The enhanced sensitivity scaling of QCM results essentially from the many-body entanglement in the ground state at the critical points. The spontaneous symmetry breaking in the formation of long-range order means the many-body entanglement is fragile to local measurement or coupling to local environments, which infers that any sensitivity scaling beyond the standard quantum limit would be diminished by local decoherence.


\section{Conclusions} Using the one-dimensional transverse-field Ising model coupled to local bosonic environments as a representative example, we find that the local decoherence reduces the scaling of the sensitivity of quantum critical metrology from the sub-Heisenberg limit to the standard quantum limit. Such reduction is understood by the picture that the coupling to the local environments amounts to a local measurement of the spins, which causes a globally entangled state collapse into a product state. Using universal scaling laws we demonstrate that the conclusion should hold for general quantum phase transitions that have a $\phi^{4}$ low-energy theory. Since the symmetry-breaking quantum phase transitions are in general associated with the macroscopic superposition of short-range entangled states (such as product states) \cite{Beekman2019}, the diverging susceptibility at the quantum criticality is inevitably associated with the fragility of the cat states in noisy environments.

It is intriguing to ask whether quantum criticality that does not involve symmetry breaking (e.g., those due to the formation of topological orders \cite{Zeng2019}) could offer QCM robust against local decoherence \cite{Bartlett2017} since the topological cat states are macroscopic superposition of locally indistinguishable, long-range entangled states and are therefore immune to local perturbations.  However, the insensitivity to local noises of topological cat states means insensitivity to local parameters (such a field coupled uniformly to individual particles). It remains an open, interesting question whether and how a measurement of a non-local parameter (which would require quantum measurement on a non-local basis) could be designed to exploit the local-decoherence-resilience of topological quantum criticality.

\textit{Acknowledgements.} This work was supported by RGC/NSFC Joint Research Scheme Project N\_CUHK403/16.

\appendix 

\section{Phase diagram by Monte Carlo simulation}
The phase diagram and the correlation function of the Ising action in Eq. (5) of main text are obtained from the Monte Carlo simulation. An example is the average of an observable $\langle O\left(\left\{s_{x,y}\right\}\right)\rangle$ as
{\small
\begin{align}
\left\langle O\left({{s}_{x,y}}\right)\right\rangle=&\sum_{\{s_{i,\tau}\}}{P_\mathrm{eq}\left(\left\{s_{i,\tau}\right\}\right)}O\left(\left\{s_{i,\tau}\right\}\right) \approx L^{-1}\ \sum_{l=1}^{L}O\left(\left\{s_{x,y}^{\left(l\right)}\right\}\right),
\label{eq:eq:ImportanceSampling}
\end{align}}
where $P_\mathrm{eq}\left(\left\{s_{i,\tau}\right\}\right)=\frac{1}{Z}e^{-S\left(\left\{s_{i,\tau}\right\}\right)}$ is the equilibrium distribution for the action $S\left(\{s_{i,\tau}\}\right), \{s_{x,y}^{\left(l\right)}\}$ is the $l$-th spin configuration sampling according to the probability distribution  $P_\mathrm{eq}\left(\left\{s_{i,\tau}\right\}\right)$ , and $L$  the total number of sampled configurations. When L\ is large enough, the sampling result becomes exact.

The sampling with a given probability distribution $P_\mathrm{eq}\left(\left\{s_{i,\tau}\right\}\right)$ is generated from the Metropolis algorithm, in which the $(l+1)$-th spin configuration $\left\{s_{i,\tau}^{\left(l+1\right)}\right\}$  is generated from the $l$-th  configuration $\left\{s_{i,\tau}^{\left(l\right)}\right\}$  by a stochastic walk with an acceptance probability $P\left(\left\{s_{i,\tau}^{\left(l+1\right)}\right\}\middle|\left\{s_{i,\tau}^{\left(l\right)}\right\}\right)=
\frac{P_\mathrm{eq}\left(\left\{s_{i,\tau}^{\left(l+1\right)}\right\}\right)}{P_\mathrm{eq}\left(\left\{s_{i,\tau}^{\left(l\right)}\right\}\right)}$. The proper choice of the conditional probability makes the $(l+1)$-th spin-configuration sampling satisfies the same equilibrium distribution $P_\mathrm{eq}$ as that of the $l$-th spin-configuration sampling.

The initial spin configuration sampling that satisfies $P_\mathrm{eq}$ is also generated by similar stochastic walks, in which, starting from an arbitrary, random spin configuration, after $m \gg 1$ steps of stochastic walks, the probability distribution of the spin configuration $\left\{s_{i,\tau}^{\left(m\right)}\right\}$ converges to the equilibrium distribution $P_\mathrm{eq}$, as a result of the detailed balance condition.

\begin{figure}[htbp]
   \centering
   \includegraphics[width=0.5\textwidth]{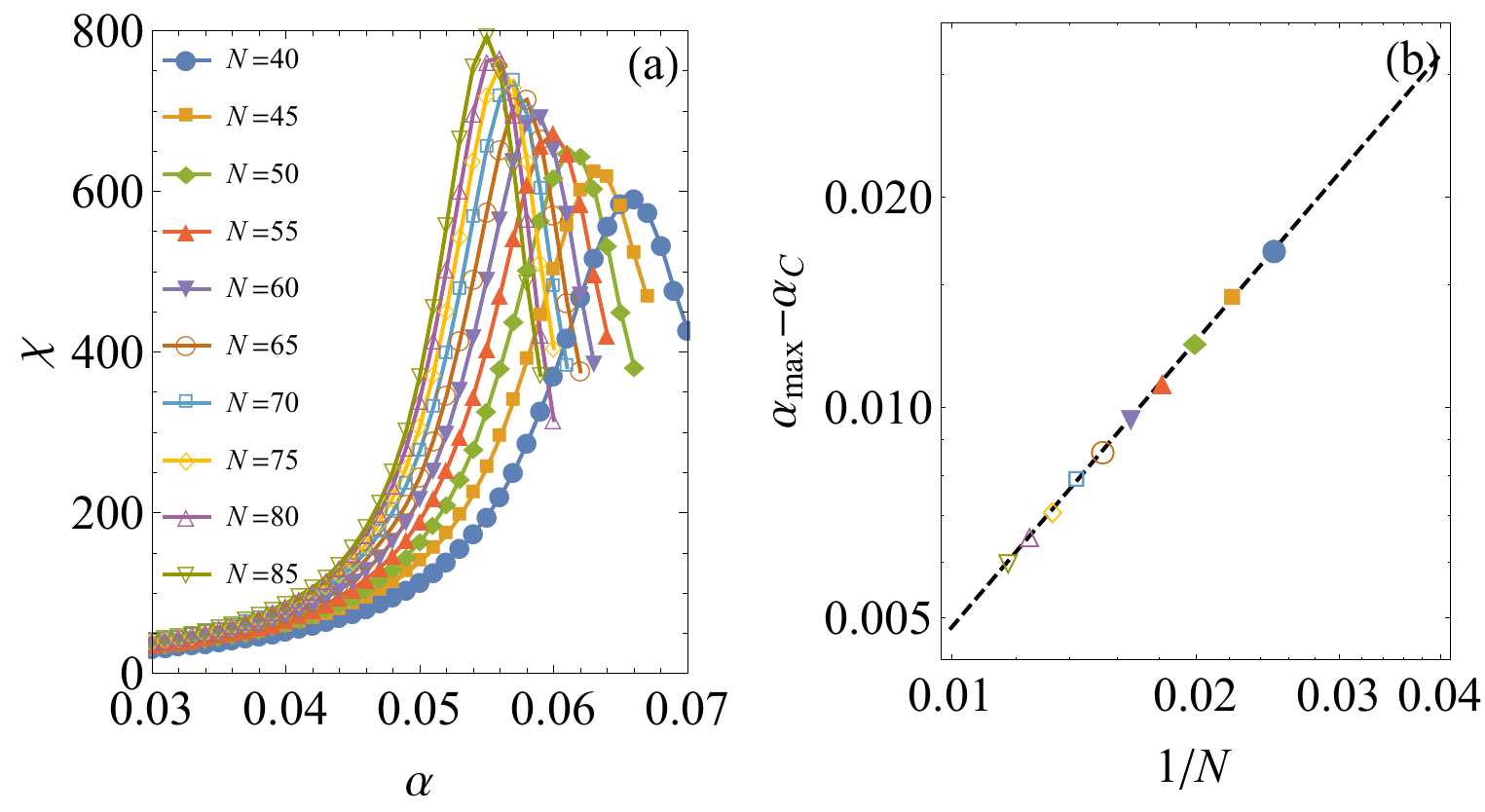}
   \caption{Magnetic susceptibilities as functions of the noise strength for different system sizes $N$ . Here we fix  $B=1.0$,  $J/J_C=0.8$, and  $N_\tau=500$.  (b)  Scaling of the peak position $\alpha_{\rm{max}}$ for the magnetic susceptibility with system size $1/N$. The dashed line is the fitting $\alpha_{\rm{max}}=\alpha_C+a\ N ^{-1/\nu_a}$, with the fitting parameters $\alpha_C=0.049$ and $\nu_\alpha=0.73$.}
   \label{fig:Susceptibility}
 \end{figure}

For each fixed value of  $J/J_C$, we gradually scan the noise strength $\alpha$ to find the critical point $\alpha_C$.  The magnetic susceptibility, $\chi =\frac{\partial M}{\partial h}=\frac{\left\langle \left(\sum_{i,\tau} s_{i,\tau}\right)^2\right\rangle-\left\langle \left(\sum_{i,\tau} s_{i,\tau}\right) \right\rangle^2}{N N_\tau}$, peaks at the noise strength $\alpha_{\rm{max}}$, whose value depends on the system size $N$  (see Fig. \ref{fig:Susceptibility}(a)). Here,  $N_\tau$ is chosen to be a large enough number. According to the finite-size scaling hypothesis,  $|\alpha_{\rm{max}}-\alpha_C|$ scales with $N$  by
\begin{equation}
	\left|\alpha_{\rm{max}}-\alpha_C\right|\sim N^{-1/\nu_\alpha}.
\label{eq:critial-ac}
\end{equation}
A numerical fitting $\alpha_{\rm{max}}=\alpha_C+a N^{-1/\nu_a}$ yields the critical point $\alpha_C$ and the critical exponent $\nu_\alpha$
 (see Fig. \ref{fig:Susceptibility} (b)).

\begin{figure}
   \centering
   \includegraphics[width=0.5\textwidth]{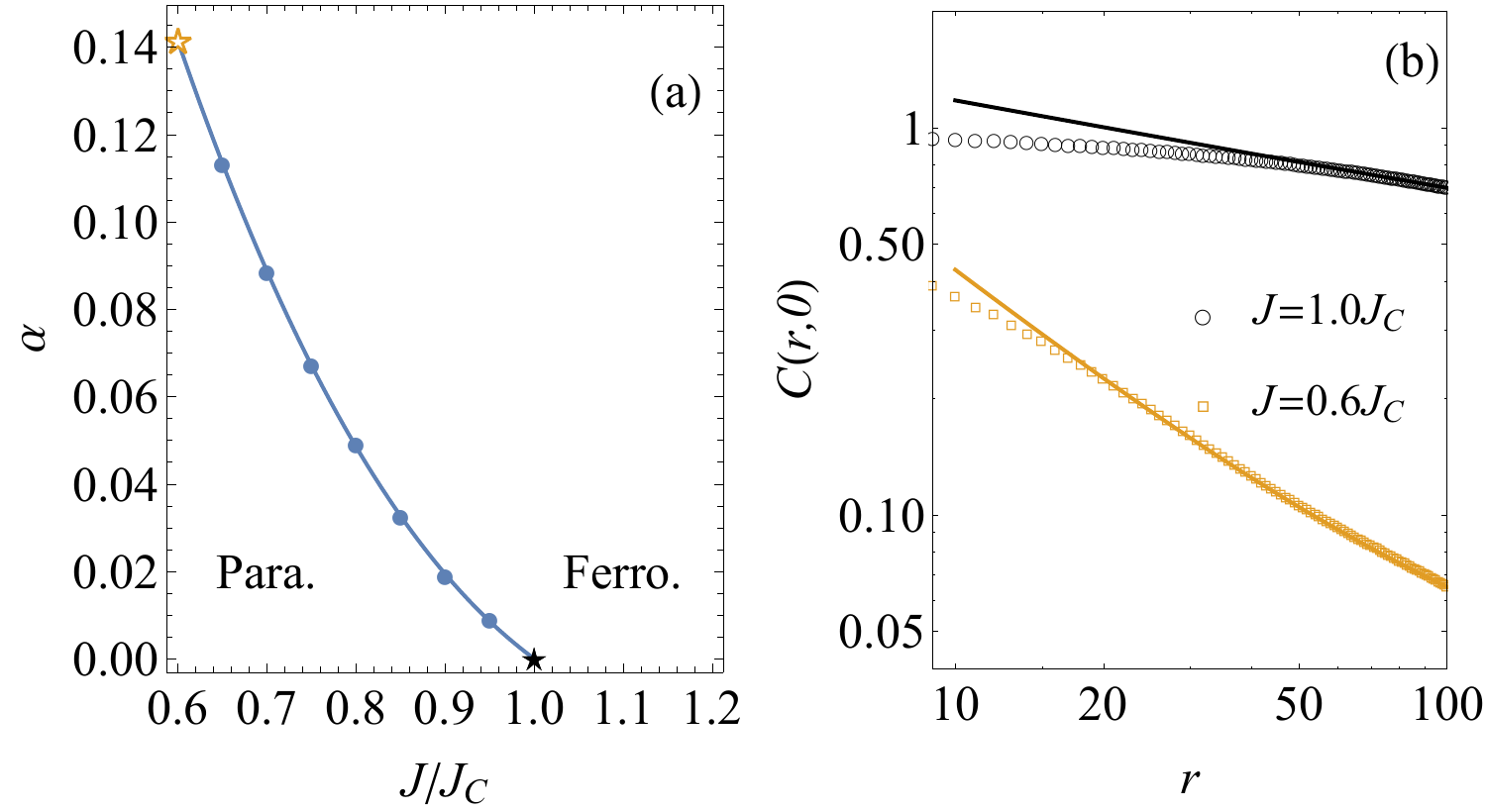}
   \caption{Critical behaviors of the one-dimensional transverse-field Ising model coupled to local environments. (a) The phase diagram for quantum phase transitions between the paramagnetic phase (Para.) and the ferromagnetic (Ferro.) phases with the spin coupling $J$ and the coupling strength $\alpha$ to local environments. (b) The equal-time correlation function $C\left(r,0\right)=\langle \hat{\sigma}^{z}_{i+r} \hat{\sigma}^{z}_{i}\rangle-\langle \hat{\sigma}^{z}_{i} \rangle^2$ for two points on the critical line, $(\frac{J}{JC}, \alpha_{C})=(1,0)$ and $(0.6, 0.141)$ (marked by the solid and open stars in (a), respectively). The lines are fitting $C(r,0)=ar^{-b}+c$ with $b=0.25(2)$ for $J=J_C$ and $\alpha_C=0.0$ (black) and $b=1.0(2)$, for $J=0.6J_C$ and $\alpha_C=0.141$ (orange). In the numerical calculation, we take the system size $(N ,N_\tau)=(200, 3000)$.}
   \label{fig:phaseDiagram}
 \end{figure}

 The critical noise strength $\alpha_C$ as a function of the ferromagnetic coupling strength defines the phase diagram, shown in Fig. \ref{fig:phaseDiagram} (a). In Fig. \ref{fig:phaseDiagram} (b), we compare the critical correlation function between the noise-free quantum critical point ($J =J_C$  and $\alpha_C=0$) and the noisy one ($J =0.6J_C$ and $\alpha_C=1.141$). At the critical point, the scaling hypothesis dictates a scaling relation $C\left(r\gg1,0\right)\sim r^{-(z+\eta-1)}$. The numerical fitting $C\left(r,0\right)=a r^b+c$ yields the critical exponent $z +\eta-1 = 0.25(2)$ for the the noise free critical point and $z+\eta-1=1.0(2)$ for the noisy one. With the noise the correlation decays faster, indicating that the many-body correlation or entanglement is weaker.

\section{Renormalization group analysis}
\begin{figure}[tpbh]
  \centering
  \begin{tikzpicture}[scale=0.2]
    \node at (-19,1.5) {(a)};
    \coordinate (o1) at (-16.5,0);
    \coordinate (o2) at (-13.5,0);
    \coordinate (o3) at (-10.5,0);
    \draw[black, thick] (o1) to (o3);
    \node[fill=black, shape=circle, inner sep=1.0pt] at (o2) {};
    \draw[thick, black,scale=10] (o2) to [in=50, out=130, loop] (o2);

    \node at (-6.5,1.5) {(b)};
    \coordinate (a1) at (-3,1);
    \coordinate (a2) at (3,1);
    \coordinate (b1) at (-4.5,1.5);
    \coordinate (b2) at (-4.5,0.5);
    \coordinate (c1) at (4.5,1.5);
    \coordinate (c2) at (4.5,0.5);
    \node[fill=black, shape=circle, inner sep=1.0pt] at (a1) {};
    \node[fill=black, shape=circle, inner sep=1.0pt] at (a2) {};
    \draw[black,thick] (a1) to [bend right= 70](a2);
    \draw[black,thick] (a1) to [bend left = 70](a2);
    \draw[black,thick] (b1)--(a1);
    \draw[black,thick] (b2)--(a1);
    \draw[black,thick] (c1)--(a2);
    \draw[black,thick] (c2)--(a2);

    \node at (8.5,1.5) {(c)};
    \coordinate (d0) at (10,1);
    \coordinate (d1) at (12.5,1);
    \coordinate (d2) at (14.5,1);
    \coordinate (d3) at (16.5,1);
    \coordinate (d4) at (19,1);
    \draw[black,thick] (d2) circle (2);
    \draw[black,thick] (d0) to (d4);
    \node[fill=black, shape=circle, inner sep=1.0pt] at (d1) {};
    \node[fill=black, shape=circle, inner sep=1.0pt] at (d3) {};
    \end{tikzpicture}
    \caption{Leading order diagrams of the $\phi^4$ action for the renormalization group analysis.}
  \label{fig:FeynmanDiagram}
\end{figure}
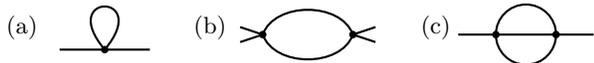
The low-energy effective field theory for the Ising model coupled to local bosonic baths is captured by the $\phi^4$ action
\begin{eqnarray}
	S&=&\iint dx^{D}  d\tau \left[\frac{1}{2}(\bm{\nabla}_{x}\phi(\mathbf{x},\tau))^{2} +\frac{1}{2}(\partial_{\tau} \phi(\mathbf{x},\tau))^{2}+\frac{\tilde{\Delta}}{2} \phi^{2}(\mathbf{x},\tau)  \right. \nonumber\\
	 && \left. -A \int d\tau^{\prime} \frac{\phi(\mathbf{x},\tau)\phi(\mathbf{x},\tau^{\prime})}{2\pi(\tau-\tau^{\prime})^{2}} +\frac{\mu_{0}}{4!} \phi^{4}(\mathbf{x},\tau)\right].
  \label{eq:phi4}
\end{eqnarray}
From the dimensional analysis, we get  $\left[\phi\left(\mathbf{x},t\right)\right]\sim\xi^\frac{2-D-z}{2}$  and the upper-critial dimension $D=4-z$ with $z=2$ for the interaction-free theory.  Above the upper-critical dimension, i.e., $D\geq2$, the $\phi^4$ term becomes irrelevant and the renormalization of the system flows to the interaction-free theory.  Around the upper-critical dimension, an expansion with respect to $\epsilon=2-D$ is used to analyze the effect of the $\phi^4$ term. Following the renormalization group analysis~\cite{Pankov2004}, we obtain the flow equation that includes the leading order loop diagrams (shown in Fig. \ref{fig:FeynmanDiagram})
\begin{subequations}
\label{eq:RGflow}
  \begin{align}
  \lambda\frac{d }{d\lambda} \tilde{\Delta}(\lambda)=& 2 \tilde{\Delta} + \frac{g}{2} -\frac{\tilde{\Delta} g}{2},\label{eq:RG-flow-r} \\
  \lambda\frac{d }{d\lambda} g(\lambda)=& \epsilon g - \frac{3}{2}g^2, \label{eq:RG-flow-g} \\
  \lambda\frac{d }{d\lambda} \ln(\tilde{C}^{-1}(\mathbf{q},0))=& 2-\frac{12-\pi^{2}}{48} g^{2}, \label{eq:RG-flow-cq} \\
  \tilde{C}^{-1}(0,i\omega)\approx & (1+c g^{2})A|\omega|,\label{eq:RG-flow-co}
\end{align}
\end{subequations}
where $\lambda$ is the scaling factor of the scaling transformation $q \rightarrow q^\prime= \lambda q$, $g\left(\lambda\right)=\frac{2\mu_0\left(\lambda\right)} {\pi\Gamma\left(\frac{D}{2}\right) (4\pi)^\frac{D}{2}}$ is the renormalized interaction strength,and $\tilde{C}\left(\mathbf{q},i\omega\right)=\langle\phi\left(-\mathbf{q},i\omega\right)\phi\left(\mathbf{q},-i\omega\right)\rangle$ is the correlation function in Fourier space. The fixed point is $\tilde{\Delta}_\ast\approx-\frac{\epsilon}{6}$ and $g_\ast=\frac{2}{3}\epsilon$. Around the critical point, the scaling analysis yields $[\tilde{C}(\mathbf{q},0)]\sim \xi^{2-\eta}$  with $\eta=\frac{12-\pi^2}{108}\epsilon^2$ and $[\tilde{\Delta}-\tilde{\Delta}_{\ast}]\sim \xi^{-\frac{1}{\nu}}$ with $\nu=\frac{1}{2}+\frac{\epsilon}{12}$. The linear dependence on $|\omega|$ of $\tilde{C}(\mathbf{q},i\omega)$  indicates that the susceptibility has a universal expression $\tilde{C}(\mathbf{q},i\omega)=q^{2-\eta} \phi(\frac{\omega}{c q^{2-\eta}})$, which leads to the universal scaling relation $[\omega]\sim q^{2-\eta}$ and therefore the scaling law for the quantum phase transition with the decoherence effects
\begin{equation}
z+\eta=2.
\label{eq:ScalingLaw-App}
\end{equation}
\bibliographystyle{apsrev4-1}
\bibliography{QCM}
\end{document}